\documentclass[preprint,11pt,authoryear]{elsarticle}

\usepackage[margin=1in]{geometry}
\usepackage{amsmath,amsfonts,amssymb,amsthm}
\usepackage{mathtools}
\usepackage[dvipsnames]{xcolor}
\usepackage{hyperref}
\usepackage{derivative}
\usepackage{mathrsfs}
\usepackage{booktabs}
\usepackage{float}
\usepackage{subcaption}
\usepackage{enumitem}

\usepackage{fancyhdr}
\newcommand{\x}{\mathbf{x}}
\newcommand{\y}{\mathbf{y}}
\newcommand{\Y}{\mathbf{Y}}

\renewcommand{\v}[0]{\mathbf{v}}
\newcommand{\w}[0]{\mathbf{w}}
\renewcommand{\u}[0]{\mathbf{u}}

\newcommand{\h}{\mathbf{h}}
\newcommand{\p}{\mathbf{p}}

\newcommand{\R}{\mathbb{R}}

\renewcommand{\S}{{\mathbf{S}}}
\newcommand{\C}{\mathbf{C}}
\newcommand{\Ac}{\mathcal{A}}
\newcommand{\Bc}{\mathcal{B}}
\newcommand{\Cc}{\mathcal{C}}

\newcommand{\Ascr}[0]{\mathscr{A}}
\newcommand{\Bscr}[0]{\mathscr{B}}
\newcommand{\Cscr}[0]{\mathscr{C}}

\newcommand{\Fc}{\mathcal{F}}

\renewcommand{\Pr}{\mathbb{P}}
\newcommand{\E}[0]{\mathbb{E}}
\newcommand{\normal}[0]{\mathcal{N}}

\newcommand{\Zero}[0]{\mathbf{0}}

\newcommand{\mbf}[1]{{\mathbf{#1}}}
\newcommand{\bs}[1]{{\boldsymbol{#1}}}
\newcommand{\BETA}[0]{{\boldsymbol{\beta}}}
\newcommand{\MU}[0]{{\boldsymbol{\mu}}}
\newcommand{\XI}[0]{{\boldsymbol{\xi}}}
\newcommand{\SIGma}[0]{{\boldsymbol{\sigma}}}
\newcommand{\SIGMA}[0]{{\boldsymbol{\Sigma}}}
\newcommand{\THETA}[0]{{\boldsymbol{\theta}}}
\newcommand{\DELTA}[0]{{\boldsymbol{\delta}}}

\newcommand{\OMEGA}[0]{\boldsymbol{\omega}}

\newcommand{\PSI}[0]{{\boldsymbol{\psi}}}
\newcommand{\ZETA}[0]{{\boldsymbol{\zeta}}}

\newcommand{\GammaFunc}[0]{\operatorname{Gamma}}

\newcommand{\tr}[0]{\operatorname{tr}}
\newcommand{\diag}[0]{\operatorname{diag}}

\newtheorem{theorem}{Theorem}[section]
\newtheorem{lemma}[theorem]{Lemma}

\theoremstyle{definition}
\newtheorem{setting}[theorem]{Setting}
\newtheorem{assumption}[theorem]{Assumption}

\theoremstyle{remark}
\newtheorem{remark}[theorem]{Remark}

\newcommand{\variance}[0]{{\operatorname{Var}}}

\newcommand{\valueatrisk}[0]{{\operatorname{VaR}}}
\newcommand{\CFVaR}[0]{{\operatorname{CFVaR}}}

\newcommand{\Tcal}{\mathcal{T}}

\hypersetup{
    colorlinks={True},
    linkcolor={blue},
    citecolor={SpringGreen},
}

\begin{document}
\begin{frontmatter}

\title{Optimal Option Portfolios for Skew-Elliptical $t$ Returns}
\author[mcmaster]{Kyle Sung}
\author[mcmaster]{Traian A. Pirvu}
\address[mcmaster]{Department of Mathematics and Statistics, McMaster University}
\date{\today}

\begin{abstract}
    This paper explores option portfolio optimization when the underlying returns are skew-elliptical $t$-distributed. We use the variance and value at risk (VaR) to measure portfolio risk. The novelty of our work is the departure from the traditional normal returns setting, allowing investors to capture both heavy-tailed and skewed market dynamics. We provide explicit portfolio weights for the variance and VaR approximation. Our second contribution is the numerical representation of portfolio weights, obtained from numerical optimization for better VaR approximations. The effect of skewness on the portfolio weights is quantified by comparing our optimal skew $t$ weights with those generated in the Student $t$ setting. We also find that, as expected, a better VaR approximation risk measure yields optimal portfolio weights which are more different than the variance optimal weights.
\end{abstract}

\end{frontmatter}


\section{Introduction}

Options are financial derivatives that give a party the right, but not the obligation, to purchase or sell an underlying asset at a given, pre-determined price on a pre-determined date. Since an option mitigates the risk of a downside movement compared to owning the underlying asset itself, a party must pay a premium to the writer of the option to purchase an option contract. \citet{BlackScholes1973} produce a formula for the fair price of a European-style option, assuming the underlying follows a geometric Brownian motion.

Constructing an optimal portfolio of options is an ubiquitous problem in mathematical finance. Portfolio managers, financial advisors, and investors are interested in finding the optimal allocation of investments which maximize their returns and minimize their risk. As such, portfolio optimization on a variety of different asset portfolios has been explored~\citep{Maheshwari2020,Metel2017,Pourbabaee2016}.

Various works in the literature have explored finding the optimal number of units of each option to buy to minimize the risk of a portfolio's loss. Sometimes, the variance of portfolio loss is used as a risk measurement~\citep{Glasserman2003} or in portfolio optimization~\citep{Jewell2013}, however, variance is not an ideal risk measurement, as variance penalizes upside volatility and  downside volatility equally. An alternative risk measure, the Value at Risk (VaR), measures the potential losses a portfolio could experience for any arbitrary confidence level. Recent work by \citet{Cesarone2023} has considered a risk-reward tradeoff in the form of the mean-VaR portfolio optimization problem. \citet{Pang2023} investigate the portfolio selection problem against systemic risk. 

Since the VaR is defined as a percentile, there is no analytic expression unless the distribution of the portfolio loss is known. This is true even in the simple case of the underlying returns being Gaussian due to the nonlinearity of the option price formula. 

One solution is to use the delta-gamma approximation, the second-order Taylor expansion of the portfolio loss~\citep{Siven2009}. \citet{CUI20132124} use the delta-gamma approximation, then employ the Cornish-Fisher expansion to find an approximation for the VaR. \citet{Pan2023} produces explicit expressions for the optimal weights for an option portfolio to minimize the variance and Cornish-Fisher expansion of the VaR when assuming the underlying stock returns are normally distributed. This is a common assumption: other recent works in this area include the follow up by \citet{Chen2013} on work by \citet{CUI20132124}, and assumes multivariate mixture of normal distributions for asset returns. \citet{Zhu2020} introduce the risk of crash into portfolio management. 

The financial markets often exhibit behaviours which cannot be well described by a normal distribution, which has thin tails that underestimate the probability of large movements~\citep{Glasserman2003}. Thus, the Student $t$-distribution is often used to describe the underlying returns~\citep{basnarkov2019optionpricingheavytaileddistributions,Cassidy_2010} due to its heavier tails. As such,
option portfolio optimization when the underlying returns are Student $t$-distributed is more realistic in spite of some option pricing complications arising within this paradigm.

\citet{Cassidy_2010,Cassidy01082013} produce a fair price for European options written on underlying assets with returns described by a log Student $t$-distribution. \citet{Hu01012010} perform data-driven and numerical portfolio optimization assuming that the return of the underlying stocks are Student $t$-distributed and skewed $t$-distributed. 

Financial markets often exhibit behaviours which are not symmetric. Many studies have noted that, in addition to having heavier tails, markets tend to exhibit larger crashes than booms \citep{Echaust20142234,Cotter01102006}.
Though using the $t$-distribution instead of the normal distribution helps address the heavier tails, being symmetric about its mean, the $t$-distribution places equal probability on upward and downward movements. Thus, one may be interested in a generalization to model asymmetry in price movements common for financial securities. As a result, much research on elliptical distributions has focused on studying families of skewed distributions \citep{AzzaliniCapitanio2003Skewt,YinBalakrishnan2024SkewElliptical,AzzaliniCapitanio1999SkewNormal,Wang2024SkewT,KIM2003417,DemartaMcNeil2005tCopula,genton2005generalized,GENTON2001319,BRANCO200199,Lin2007RobustSkewT}. 

Recently, skew-elliptical distributions have received a lot of attention and are used in various fields such as Bayesian inference \citep{Anceschi03042023}, actuarial sciences \citep{ELING2012239,BOLANCE2008386,VERNIC2006413}, risk theory \citep{Shushi2017SkewElliptical}, and capital markets data \citep{ELING201445}.
Based on statistical tests, \citet{ELING201445} shows that the skew-elliptical $t$-distribution is a better fit to the capital markets data
considering a diversified portfolio spanning several asset classes, showing that the skew $t$ has approximately 40\% improvement in Kolmogorov-Smirnov goodness of fit; the Student-t captures fat tails but forces symmetric risk, while skew-elliptical models capture the empirically observed asymmetry between gains and losses without sacrificing tail realism.

Though it is known that the skew-elliptical $t$-distribution better describes the behaviour of the underlying returns in the financial markets, to the best of our knowledge, the general question of the optimal options portfolio in this setting remains open.

\paragraph{Contribution} Our contribution is to analyze and solve option portfolio optimization problems when the underlying returns are skew $t$-distributed. We produce explicit expressions for the optimal portfolios under minimization of the variance and two-term Cornish-Fisher VaR approximation.
Our numerical results indicate that the optimal weights for the three-term expansion diverge from the variance weights more significantly than those of the two-term expansion only in the case of the skew $t$ distribution, suggesting that constructing robust portfolios necessitates accounting for skewed distributions. This effect, however, does not hold for the standard $t$ distribution. Additionally, the portfolio outlook may shift under the standard $t$ distribution.

\subsection{Outline}

Section~\ref{sec:model} provides an overview of the Gosset Formula, assumptions, and background needed for portfolio optimization. Section~\ref{sec:portfolio_optimization} presents our main results: optimal options portfolios under variance and VaR minimization assuming that returns follow a Student $t$-distribution. In Section~\ref{sec:estimation}, we perform parameter estimation to fit a skew $t$ distribution to financial data. Our theoretical results are used to produce examples of optimal portfolios in Section~\ref{sec:numerics}. Section~\ref{sec:conclusion} concludes. The proof of our main result and technical details on the distributions and data sets are located in the appendix.

\section{The Model} \label{sec:model}

\subsection{Notation}
All random variables are henceforth defined on the probability space $(\Omega, \Fc, \Pr)$, and expectations are taken with respect to the probability measure $\Pr$. Let $\odot$ denote the Hadamard product and let $\otimes$ denote the outer product. For a matrix $X \in \R^{M\times N}$, we let $X_{[\bullet, i]}$ denote its $i^{\text{th}}$ column.

\subsection{Setting}

\noindent We will work in the following setting.

\begin{setting}[Optimal Portfolios]
    \label{setting:optimal_portfolios}
    Consider a portfolio consisting of $M$ options written on $N$ underlying stocks. Let $x_1,\ldots, x_M$ denote the number of shares held of each option and let $\x = (x_1,\ldots,x_M)$ denote the vector of number of shares. 
    Denote by $V_m \coloneqq V_m(\S, t)$ the value of option $m$ at time $t$ with underlying price $\S$, where $\S \coloneqq (S_1,\ldots,S_N)$. Denote by
    \begin{equation}
        \label{eqn:option_prices_vector}
        \v \coloneqq \v(\S,t) = \big(V_1(\S,t),\ldots,V_M(\S,t)\big)
    \end{equation} the vector of option prices. The value of the portfolio is given by 
    \[
        V(\x;\S,t) 
        = 
        \x^\top \v(\S,t) 
        = 
        \sum_{m=1}^M x_m V_m(\S,t).
    \]
    Note that we can easily convert from a vector of shares to a vector of weights as follows:
    \[
        \w = \dfrac{\v \odot \x}{\v^\top \x}
        .
    \]

    Let $\Delta V (\x)$ denote the portfolio gain over the time interval $[t,t + \Delta t]$, which can be written explicitly as \begin{equation}
        \Delta V(\x) = V(\x; \S + \Delta \S, t + \Delta t) - V(\x; \S,t)
        .
    \end{equation}
\end{setting}

\begin{remark}[Generality of Option Portfolios]
    One may wonder about the motivation of portfolio optimization in the option space instead of
    directly on stocks. In fact, considering option portfolios allows for the more general setting. In
    particular, long and short equity positions are embedded in the option space via put-call parity. One can similarly construct long and short forward and future positions by entering
    option positions and borrowing or lending money. Thus, one can use these results to construct
    optimal portfolios for nearly any commonly-traded security. Further, our results are completely
    general, and one could replace the options in the portfolio with any other assets, so
    long as the first-order time and underlying partials and second-order underlying partials are known.
\end{remark}

s
We now formalize our main assumption---that the underlying returns are skew $t$-distributed---from which we will derive our main results of optimal option portfolios.

\begin{assumption}[Skew $t$-Distributed Returns]
    \label{assumption:nonzero_drift_t_distributed_returns}
    In Setting~\ref{setting:optimal_portfolios}, fix a location vector $\MU \in \R^M$, a scale matrix $\SIGMA \in\R^{M\times M}$, a degrees of freedom $\nu \in \R$, and skewness vector $\OMEGA \in \R^M$. We assume that $\Delta \S $ has the multivariable skew-elliptical $t$-distribution with parameters $\big(\MU, \SIGMA, \nu, \OMEGA\big)$. That is, 
    \[
        \Delta \S  \sim t_N^{\operatorname{skew}} 
        \big (
            \MU, \SIGMA, \nu, \OMEGA
        \big )
        .
    \]
\end{assumption}
\begin{remark}
    When $\bs\omega \to \Zero$, one has $t_N^{\operatorname{skew}}(\MU, \SIGMA, \nu, \bs\omega) \overset{d}{\to} t_N (\MU, \SIGMA, \nu)$. Thus, our results easily translate to the standard multivariable Student $t$-distribution.
\end{remark}

\subsection{Pricing the Option}

We proceed with a modification to the Gosset Formula~\citep{Cassidy_2010} to price European options where the underlying returns follow a log skew $t$-distribution. We call this so called price the \textit{Skew-Gosset} formula.

Let $p_p$ and $p_c$ respectively denote lower and upper truncation probabilities and let $x_p$ and $x_c$ respectively denote the upper and lower quantiles such that $p_p = \Pr (z \leq x_p) $ and $p_c = \Pr(z \leq x_c)$. Assume that the underlying asset price takes only values $x_p \leq S(t) \leq x_c$ and let $f_t$ denote the p.d.f.\ of the returns distribution. Let $f_r$ denote the (unnormalized) \emph{truncated} p.d.f.\ of the returns distribution:
\[
    f_r (z) = \begin{cases}
        0 & \text{~if~} z < x_p
        \\
        f_t(z) & \text{~if~} x_p \leq z \leq x_c
        \\
        0 &\text{~if~} z > x_c,
    \end{cases}
\]
which is defined to ensure the convergence of the integrals needed to evaluate the fair price of the Gosset option. These choices are reasonable---for sufficiently large truncation points---as extremely high prices are effectively impossible, and thus this is still consistent with real-world asset prices. 

Let $Z$ denote the normalization integral and let $A_T$ denote the average value of the underlying until time $T$:
\[
    Z = \int_{x_p}^{x_c} \dfrac{\exp(\sigma_T z) f_r(z)}{p_c - p_p} \,\mathrm{d} z
    ,
    \quad
    A_T = \dfrac{S(0) \exp(rT)}{Z}
    .
\]
The price of the European call option is then
\[
    C_T = \int_{\tfrac{\log(K_T / A_T)}{\sigma_T}}^{x_c} \dfrac{(A_T \exp(\sigma_T z ) - K_T) f_r (z)}{p_c - p_p} \,\mathrm{d}z,
\]
assuming that $\log (K_T / A_T)/\sigma_T > x_p$, and the price of the European put option is
\[
    P_T = \int_{x_p}^{\tfrac{\log(K_T/A_T)}{\sigma_T}} \dfrac{(K_T - A_T\exp(\sigma_T z))f_r(z)}{p_c - p_p} \,\mathrm{d}z .
\]

\noindent We perform numerical integration using Scipy~\citep{2020SciPy-NMeth} and we compute the Gosset Greeks with finite difference approximations of the Gosset formulas; some explicit forms have been explored in prior works~\citep{Cassidy01082013}. Let $\THETA$, $\DELTA$, and $\Gamma$ denote the usual per-option Greeks and let $\theta \coloneqq \pdv{V}{t}$.

\begin{remark}
    We note the following three special cases: (1) in the case where the skewness parameter is zero, $f_r$ collapses to the standard $t$ density, and this option price coincides exactly with the Gosset formula \citep{Cassidy_2010,Cassidy01082013}; (2) in the case where the skewness parameter is nonzero but the degrees of freedom tends to infinity, then we are in the setting of the skew normal distribution \citep{Azzalini1996MultivariateSkewNormal,Azzalini1999MultivariateSkewNormalApplications} and the option price coincides  with the skew Brownian motion option price by \citet{Zhu13082018}; and (3) in the case where the skewness parameter is zero and the degrees of freedom tends to infinity, this option price coincides exactly with the Black-Scholes formula \citep{BlackScholes1973}. In the cases of (2) and (3), the integrals are finite, and one need not set truncation points.
\end{remark}

\subsection{Delta-Gamma Approximation}

\noindent As the distribution of the portfolio gain is not directly known, we recourse to the delta gamma approximation, which is the second-order Taylor expansion of the portfolio gain:
\[
    \Delta V (\x) = (\Delta t) \theta + \DELTA^\top (\Delta \S) + \frac{1}{2} (\Delta \S)^\top \Gamma (\Delta \S).
\]
We compute the expectation and variance of $\Delta V$ by completing the square, performing an affine transformation, and using the moments of the multivariable skew-elliptical skew $t$-distribution~\citep[Theorems 2 and 3]{KIM2003417}. We have
\begin{align*}
    \E[\Delta V (\x) ] &=
        (\Delta t)\theta + \DELTA^\top \MU + \dfrac{\nu}{2(\nu - 2)} \tr(\Gamma \SIGMA)+ \frac{1}{2} \MU^\top \Gamma \MU 
        + c\MU^\top\Gamma \h + c\DELTA^\top\h
    \\
    \variance[\Delta V (\x) ] &=
            \biggl[\frac{\nu^{2}}{2(\nu-2)(\nu-4)}\biggr] \operatorname{tr}[(\Gamma\SIGMA)^{2}]
            + \biggl[\frac{\nu^{2}}{2(\nu-2)^{2}(\nu-4)}\biggr] \operatorname{tr}[\Gamma\SIGMA]^{2}
        \\&\quad\; 
            + 
            \dfrac{\nu}{\nu-2} [\Gamma\MU + \DELTA]^\top \SIGMA [\Gamma\MU + \DELTA]  
            + 
            \dfrac{2c\nu}{\nu-3} [\Gamma\MU + \DELTA]^\top \SIGMA \Gamma \h
        \\&\quad\;
            + \dfrac{c\nu}{(\nu-2)(\nu-3)}[\Gamma\MU + \DELTA]^\top \h \operatorname{tr}[\Gamma\SIGMA]
        \\&\quad\;
            - \dfrac{c\nu}{\nu-3}
            [\Gamma \MU + \DELTA]^\top \h \h^\top \Gamma\h - c^2 ([\Gamma\MU + \DELTA]^\top \h)^2
    . 
\end{align*}
We now aim to write the expectation and variance of the portfolio gain in terms of the shares vector $\x$, so we can use it as the objective of our variance and value at risk optimization. After some algebra, we recover:
\begin{equation}
    \label{eqn:expectation_variance}
    \begin{split}
    \E[\Delta V (\x) ] &= 
         \mbf u^\top \x
    \\
    \variance[\Delta V (\x) ] &= 
        \frac{1}{2}\x^\top Q \x
    ,
    \end{split}
\end{equation}
where
\begin{align*}
    \mbf u &\coloneqq \ZETA + cB \mbf h + c D \mbf h,
    \\
    Q &\coloneqq \frac{1}{2} (\tilde Q + \tilde Q^\top),
\end{align*}
and the intermediate variables are:
\begin{align*}
    \tilde{Q} &\coloneqq U + \dfrac{4c\nu}{\nu - 3}(H + E) + \dfrac{2c\nu}{(\nu - 2)(\nu - 3)} (B + D^\top)\h\p^\top 
    \\
    &\quad\; - \dfrac{2c\nu}{\nu - 3}(B + D^\top)\h\mathbf{q}^\top - 2c^2 (B + D^\top)\h\h^\top(B + D^\top)^\top
    \\
    \ZETA &\coloneqq 
        \left[ (\Delta t) \THETA + D^\top \MU + \dfrac{\nu}{2(\nu - 2)} \p + \XI  \right]
    \\
    U &\coloneqq 
        \left[\dfrac{2\nu}{\nu -2} ((D^\top + B)\SIGMA(D^\top + B)^\top) + \dfrac{\nu^2}{(\nu - 2) (\nu -4)} R + \dfrac{\nu^2}{(\nu-2)^2(\nu-4)}\p\p^\top\right]
\end{align*}

\[
    \begin{aligned}
        \p &\coloneqq (p_1,\ldots,p_M)  &  \quad p_m &\coloneqq \tr[\Gamma^{[m]} \SIGMA ]
        \\
        D &\coloneqq (\delta_n^m)  &  \quad \delta_n^m &\coloneqq \pdv{V_m}{S_n}
        \\
        R &\coloneqq [r_{i,j}] & \quad r_{i,j} &\coloneqq \tr[\Gamma^{[i]}\SIGMA \Gamma^{[j]} \SIGMA]
        \\
        \XI &\coloneqq (\xi_1,\ldots,\xi_M)  & \quad \xi_m &\coloneqq \dfrac{1}{2} \sum\limits_{i=1}^N \sum\limits_{j=1}^N \mu_i \mu_j {\gamma}^{[i,j]}_m 
        \\
                B &\coloneqq \begin{bmatrix}
                \text{---} \hspace{-0.25cm} & \MU^\top \Gamma^{[1]} & \hspace{-0.25cm} \text{---}
                \\
                \text{---} \hspace{-0.25cm} & \vdots& \hspace{-0.25cm} \text{---}
                \\
                \text{---} \hspace{-0.25cm} & \MU^\top \Gamma^{[M]} & \hspace{-0.25cm} \text{---}
            \end{bmatrix}
        .
    \end{aligned}
\]

\subsection{Variance and VaR Risk Measures}

\noindent We use variance and VaR to quantify the portfolio risk. Variance comes with the advantage that portfolio optimization is more tractable, but its drawback is that it equally penalizes both gains and losses. One way out of this predicament is to use VaR, which penalizes only downward volatility. 

One downside is that VaR cannot be computed explicitly unless one knows the distribution of the portfolio gains or losses. The compromise is to use the Cornish-Fisher (CF) quantile approximation of the VaR. Due to the nonlinear effects, we use only the two-term and three-term quantile approximations, henceforth referred to as $\CFVaR_2$ and $\CFVaR_3$, defined for a tail risk threshold $\alpha >0$ as:
\[
    \begin{split}
        \CFVaR_2^\alpha [\Delta V(\x)] &\coloneqq 
            -\E[\Delta V(\x)] - \normal^{-1}(\alpha) \sqrt{\variance[\Delta V(\x)]}
        \\
        \CFVaR_3^\alpha [\Delta V(\x)] &\coloneqq 
            -\E[\Delta V(\x)] - \normal^{-1}(\alpha) \sqrt{\variance[\Delta V(\x)]} 
            - \frac{\left( [\normal^{-1}(\alpha)]^2 - 1 \right)}{6} \frac{ \kappa_3(\x)}{\variance[\Delta V(\x)]}
        ,
    \end{split}
\]
where $\kappa_3(\x) \coloneqq \E[(\Delta V(\x) - \E[\Delta V(\x)])^3]$ denotes the third central moment of the portfolio gain and where $\normal^{-1}(\,\cdot\,)$ denotes the inverse c.d.f.\ of the standard normal distribution.

\begin{lemma}[$\CFVaR_2$ and $\CFVaR_3$ of Portfolio Gain]
    In Setting~\ref{setting:optimal_portfolios} and under Assumption~\ref{assumption:nonzero_drift_t_distributed_returns},
    let $\alpha <1/2$ be a tail risk. If $\nu >6$, then both $\CFVaR_2$ and $\CFVaR_3$ exist and are given by:
    \begin{align*}
        \CFVaR_2^\alpha[\Delta V(\x)] 
        &=
        - \u^\top \x - \normal^{-1}(\alpha) \sqrt{\frac{1}{2}\x^\top Q\x}
        \\
        \CFVaR_3^\alpha[\Delta V(\x)] 
        &=
        - \u^\top \x - \normal^{-1}(\alpha) \sqrt{\frac{1}{2}\x^\top Q\x}
        -
        \frac{1}{6} ([\normal^{-1}(\alpha)]^2 - 1)
        \dfrac{\kappa_3(\x)}{\frac{1}{2}\x^\top Q\x},
    \end{align*}
    where
    \begin{align*}
        \kappa_3(\x) 
        &= 
        \dfrac{2\nu^3}{(\nu - 2)^3(\nu - 4)(\nu - 6)} (\x^\top \p)^3 
        +
        \dfrac{3\nu^3}{(\nu-2)^2 (\nu - 4)(\nu - 6)}(\x^\top \p)(\x^\top R\x)
        \\&\quad\;
        +
        \dfrac{3\nu^2}{(\nu - 2)^2(\nu - 4)} (\x^\top \p) [\x^\top (D^\top + B)^\top \SIGMA (D + B^\top )\x]
        + \langle \Tcal, \x\otimes \x\otimes \x\rangle,
    \end{align*}
    and where $
        \Tcal \coloneqq [\tau_{i,j,k}] \in \R^{M\times M\times M}
    $ is a tensor with entry:
    \[
        \tau_{i,j,k}
        =
        \dfrac{3\nu^2}{(\nu - 2)(\nu - 4)}
        (D_{(:,i)} + \Gamma^{[i]}\MU)^\top (\SIGMA\Gamma^{[k]}\SIGMA) (D_{(:,j)} + \Gamma^{[j]}\MU)
        +
        \dfrac{\nu^3 \tr[\Gamma^{[i]}\SIGMA\Gamma^{[j]}\SIGMA\Gamma^{[k]} \SIGMA]}{(\nu - 2)(\nu - 4)(\nu - 6)}
        .
    \]
\end{lemma}
\begin{remark}
    We require that $\nu >6$ to ensure the existence of the third moment. For the existence of only $\CFVaR_2$, $\nu >4$ is sufficient.
\end{remark}

In general, closed-form optimization of the $\CFVaR_3$ is intractable. Thus, we consider $\CFVaR_2$ for analytic results and use $\CFVaR_3$ for numerical optimization.

\section{Portfolio Optimization} \label{sec:portfolio_optimization}

\noindent We are now ready to formulate our main objectives.

\subsection{Minimizing Variance}

\noindent The minimum variance problem is to find a portfolio allocation $\x$ which solves the quadratic program
\begin{equation}
    \tag{P1}
    \label{eqn:expectation_variance:minimization}
    \begin{cases}
        \min \limits_{\x\in \R^M} \{\variance[\Delta V (\x)]\}
        \\[10pt]
        \x^\top \v(\S, t) = 1.
\end{cases}
\end{equation}
Indeed, it is well known that the solution to this optimization problem is
\begin{equation}
    \x_{\variance}^\star = \dfrac{1}{\v^\top Q^{-1} \v} Q^{-1} \v,
\end{equation}
where $Q$ was defined below Equation~\eqref{eqn:expectation_variance} and where $\v$ was defined in Equation~\eqref{eqn:option_prices_vector}.

\subsection{Minimizing Value at Risk}

\noindent The minimum Cornish-Fisher VaR problem is to find a portfolio allocation $\x$ which solves the optimization problem
\begin{equation}
    \tag{P2}
    \label{eqn:valueatrisk:minimization}
    \begin{cases}
        \min \limits_{\x \in \R^M} \{ \CFVaR_2^\alpha [\Delta V (\x)]\}
        \\[10pt]
        \x^\top \v (\S, t) = 1.
    \end{cases}
\end{equation}

\noindent The following theorem is the main analytical result of our paper, and prescribes the optimal number of shares for VaR minimization in the $t$-distributed returns setting.

\begin{theorem}
    [Optimal $\CFVaR_2^\alpha$ Portfolios]
    \label{theorem:cfvar2_optim_t_dist_returns_nonzero_drift}
    The $\CFVaR_2^\alpha$ minimization problem \eqref{eqn:valueatrisk:minimization} is solved with the optimal portfolio number of shares: \begin{equation}
        \x_{\valueatrisk}^\star 
        = 
            Q^{-1} 
            J^\top (J Q^{-1} J^\top )^{-1} 
            \PSI^\star
        ,
    \end{equation}
    where
    \begin{equation}
        \label{eqn:J_psistar}
        J \coloneqq \begin{bmatrix}
            \text{---}\!\!&\!\! \mbf u^\top  \!\!&\!\! \text{---}
            \\
            \text{---}\!\!&\!\!  \v(\S,t)^\top  \!\!&\!\! \text{---}
        \end{bmatrix},
        \quad
        \PSI^\star \coloneqq \begin{bmatrix}
            \varepsilon^\star
            \\
            1
        \end{bmatrix},
    \end{equation}
    $\mbf u$ was defined below Equation~\eqref{eqn:expectation_variance}, and $\varepsilon^\star$ is a positive constant defined in \ref{sec:proofs}.
\end{theorem}

\subsection{Minimizing $\CFVaR_3$}

\noindent We also introduce the minimum $\CFVaR_3$ problem. First, we must write the $\CFVaR_3$ explicitly in terms of the number of shares.

\noindent Now, we introduce the minimal $\CFVaR_3$ problem
\begin{equation}
    \tag{P3}
    \label{eqn:CFVaR3:minimization}
    \begin{cases}
        \min \limits_{\x \in \R^M} \{ \CFVaR_3^\alpha [\Delta V (\x)]\}
        \\[10pt]
        \x^\top \v (\S, t) = 1.
    \end{cases}
\end{equation}
which we solve computationally but not in closed form.

\section{Parameter Estimation} \label{sec:estimation}

We must now perform parameter estimation to find the distributional parameters under the skew $t$-distribution.

Following \citet{Hu01012010}, we perform estimation using the closing prices for the same five stocks selected from the Dow Jones Industrial Average: Disney (DIS), Exxon Mobil (XON), Pfizer (PFE), Altria Group (MO), and Intel (INTC). Their source of the price data is not specified; we source our price data from \texttt{yfinance} \citep{aroussi2026yfinance}.

Following the procedure outlined in Section 4, using the data between July 1 2002 and 4 August 2005, we convert the prices to daily log-returns, and drop the first 30 days, leaving 750 samples. Then, we use a $\operatorname{GARCH}(1, 1)$ model with Gaussian innovations to remove serial return dependence: that is, we fit $\alpha_0$, $\alpha_1$ and $\beta_1$ such that the log returns $R_t$ satisfy \[
    R_t = \sigma_t Z_t,
\]
where $Z_1,\ldots$ are i.i.d.\ with $Z_t \sim \normal(0,1)$ and where \[
    \sigma_t = \alpha_0 + \alpha_1 R_{t-1}^2 + \beta_1 \sigma_{t-1}^2.
\]
We refer to $Z_t$ as the filtered return, and assume that the $Z_t$ are i.i.d.\ approximately. After obtaining the approximately i.i.d.\ filtered data, we
can estimate the multivariable density.

We perform parameter estimation using maximum likelihood estimation (MLE).
The \texttt{SN} package \citep{Azzalini2026_SN_R_package} in \texttt{R} provides most of the infrastructure for fitting to the family of multivariate skew-elliptical distributions.
This yields the following optimal parameters:

\begin{table}[ht]
\centering
\begin{tabular}{ll}
\hline
Quantity & Value \\
\hline
$\mathrm{dp.mu}$ & $-0.13113, -0.095028, -0.073402, 0.32902, -0.14546$ \\
$\mathrm{dp.omega}$ & $0.069222, 0.14039, 0.14851, -0.72981, 0.17167$ \\
$\mathrm{dp.nu}$ & $6.041$ \\
$\mathrm{dp.\Sigma}$ & $\begin{pmatrix}0.73931 & 0.3017 & 0.2904 & 0.098984 & 0.34011\\0.3017 & 0.77136 & 0.31795 & 0.15848 & 0.25785\\0.2904 & 0.31795 & 0.71838 & 0.12605 & 0.2633\\0.098984 & 0.15848 & 0.12605 & 0.62268 & 0.076389\\0.34011 & 0.25785 & 0.2633 & 0.076389 & 0.66185\end{pmatrix}$ \\
$\mathrm{cp.mean}$ & $-0.029077, -0.013345, 0.028808, -0.041695, -0.010414$ \\
$\mathrm{cp.var.cov}$ & $\begin{pmatrix}1.0948 & 0.44269 & 0.42369 & 0.18581 & 0.49466\\0.44269 & 1.1465 & 0.46697 & 0.26721 & 0.37444\\0.42369 & 0.46697 & 1.0635 & 0.22632 & 0.37981\\0.18581 & 0.26721 & 0.22632 & 0.79343 & 0.16426\\0.49466 & 0.37444 & 0.37981 & 0.16426 & 0.97118\end{pmatrix}$ \\
$\mathrm{cp.gamma1}$ & $0.096803, 0.075538, 0.098388, -0.45578, 0.13681$ \\
$\mathrm{cp.gamma2M}$ & $35.594$ \\
$\log L$ & $-4808.6$ \\
\hline
\end{tabular}
\caption{Skew-$t$ model fit.}
\end{table}
\noindent The mean vector and scale matrix for the $t$-distribution in this setting are:
\begin{align*}
    \MU &= \S \odot (\Delta t)\mathbf{r}
    \\
    \SIGMA &= (\Delta t)\diag (\S) \diag (\SIGma) \C \diag (\SIGma) \diag (\S),
\end{align*}
where $\mathbf{r}$ denotes the expected annual log returns vector and $\C$ denotes the matrix of correlations.

\section{Numerical Experiments} \label{sec:numerics}

Next, we are going to obtain a representation for the optimal portfolio weights based on the both analytic and numerical optimization.

\subsection{Analytic Optimization}

\begin{figure}[h!]
    \centering
    \includegraphics[width=0.875\linewidth]{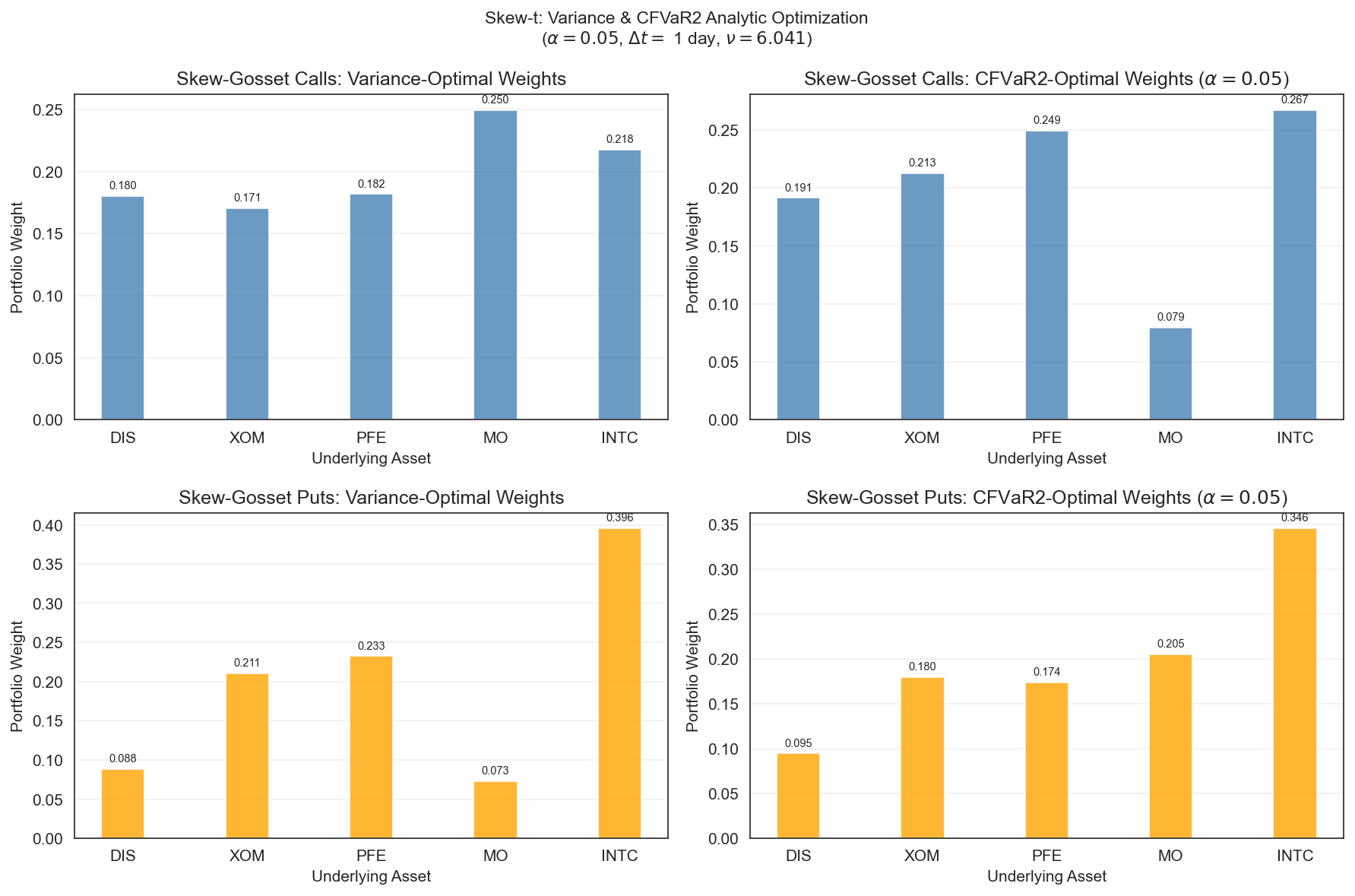}
    \caption{Optimal call and put option portfolios, under both variance and Cornish-Fisher VaR minimization, of five at-the-money options with expiry in one year, written on five stocks, where the underlying stock returns are skew $t$-distributed. Here, $\alpha = 0.05$, $\Delta t = 1/252$.}
    \label{fig:skew_analytic}
\end{figure}

We now use the results of Theorem~\ref{theorem:cfvar2_optim_t_dist_returns_nonzero_drift} which provides the optimal portfolio weights for $\CFVaR_2$ minimization in closed form. Figure~\ref{fig:skew_analytic} is produced based on this explicit solution. 

We see that the portfolio outlook is different between variance and $\CFVaR_2$. For example, the optimal weight for call options on Altria Group (MO) is $25.0\%$ under variance minimization compared with $7.9\%$ under $\CFVaR_2$ minimization.

\subsection{Numerical Optimization}

We now showcase numerical optimization of the $\CFVaR_3$ and compare it with optimization of variance and $\CFVaR_2$. Numerical multivariable optimization is performed using Scipy~\citep{2020SciPy-NMeth}. 
All scientific computation is performed using NumPy~\citep{numpy}, which provides support for array programming. In particular, we use \texttt{einsum} which offers support for tensor operations with Einstein summation, which is a standard way to deal with tensor operations.

\begin{figure}[h!]
    \centering
    \includegraphics[width=0.875\linewidth]{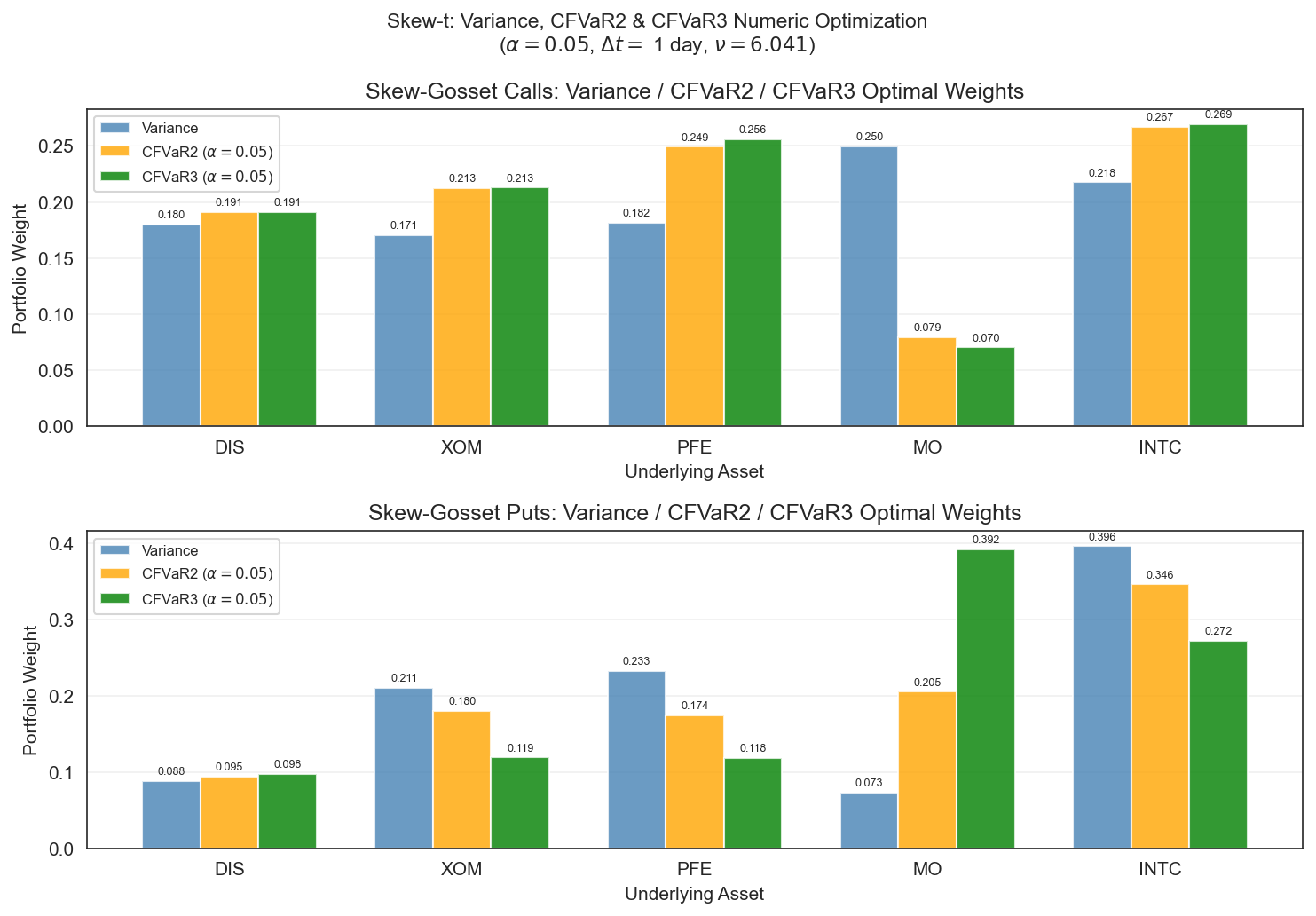}
    \caption{Optimal call (top) and put (bottom) option portfolios, under both variance (left) and Cornish-Fisher VaR (right) minimization, of five at-the-money options with expiry in one year, written on five stocks, where the underlying stock returns are skew $t$-distributed. Here, $\alpha = 0.05$, $\Delta t = 1/252$.}
    \label{fig:skew_numeric}
\end{figure}

Figure~\ref{fig:skew_numeric} displays the optimal portfolio weights for the three risk measurements: variance, $\CFVaR_2$, and $\CFVaR_3$. 

Note that the variance, $\CFVaR_2$, and $\CFVaR_3$ portfolio weights vary, for example, for call options on Altria Group (MO), the optimal variance, $\CFVaR_2$, and $\CFVaR_3$ weights are respectively $25.0\%$, $7.9\%$, and $7.0\%$. 

\subsubsection{Effect of Skewness}

We now aim to quantify the effect of skewness on the optimal portfolio weights. We do this by comparing our the skew-elliptical $t$ setting and the special case of the Student $t$ setting where $\bs\omega = \Zero$. We use a dataset of the same five stocks~\citep{Hu01012010,Pan2023}, with parameters stated in Table~\ref{table:data} for completeness.

\begin{figure}[p!]
    \centering
    \includegraphics[width=0.875\linewidth]{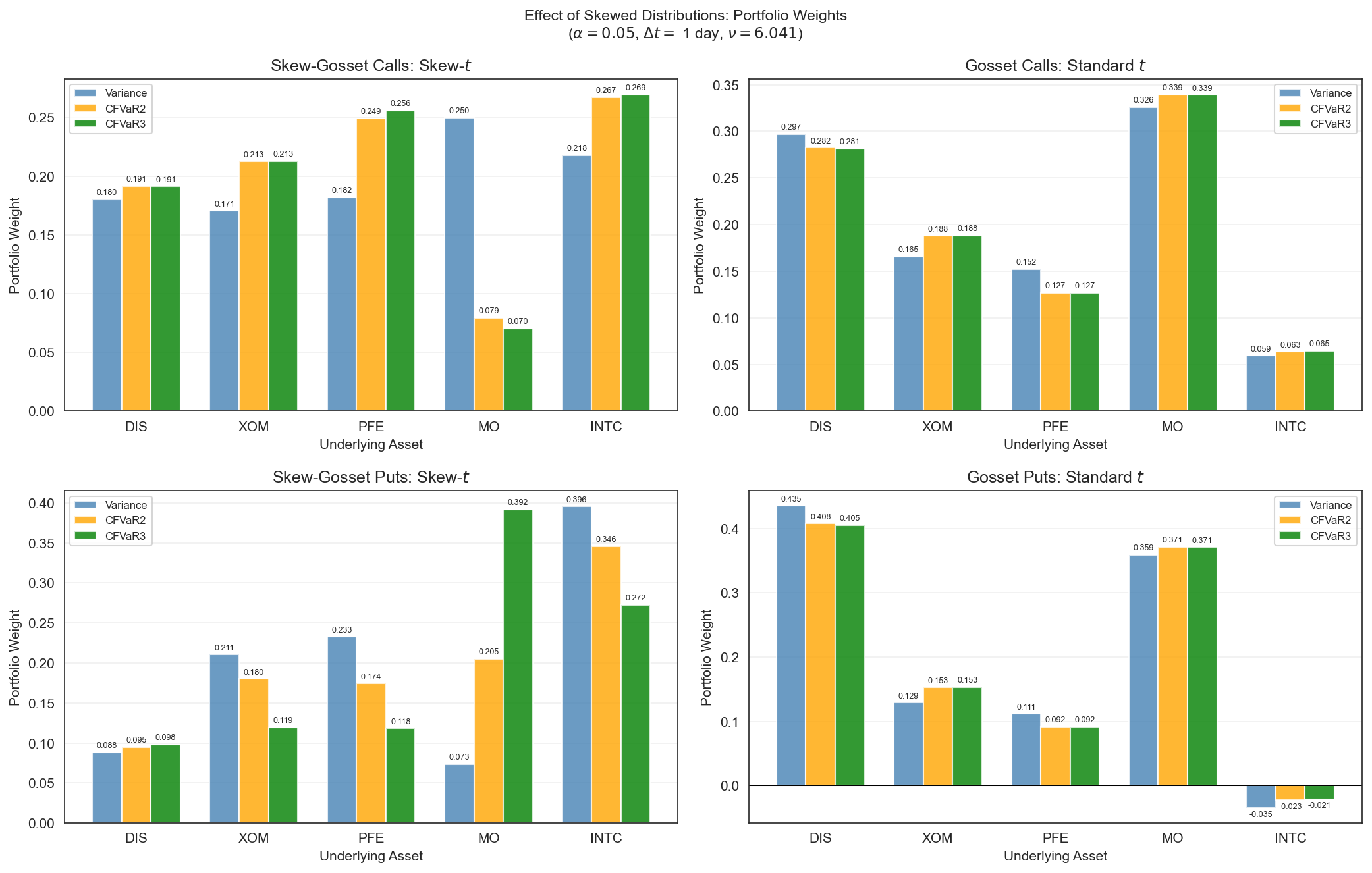}
    \caption{Optimal call (top) and put (bottom) option portfolios in the setting of skew-elliptical $t$ (left) and standard $t$ (right) distributed underlying returns. We consider objectives of minimum variance, $\CFVaR_2$, and $\CFVaR_3$. Here, $\alpha = 0.05$ and $\Delta t = 1/252$.}
    \label{fig:effect_of_skewness}
\end{figure}

Figure~\ref{fig:effect_of_skewness} displays the results.

The portfolio outlook changes when moving from the skew $t$ distribution to the standard Student $t$ distribution.

\subsubsection{\texorpdfstring{$\CFVaR_3$ vs $\CFVaR_2$}{CFVaR3 vs CFVaR2}}

\begin{figure}[p!]
    \centering
    \includegraphics[width=0.875\linewidth]{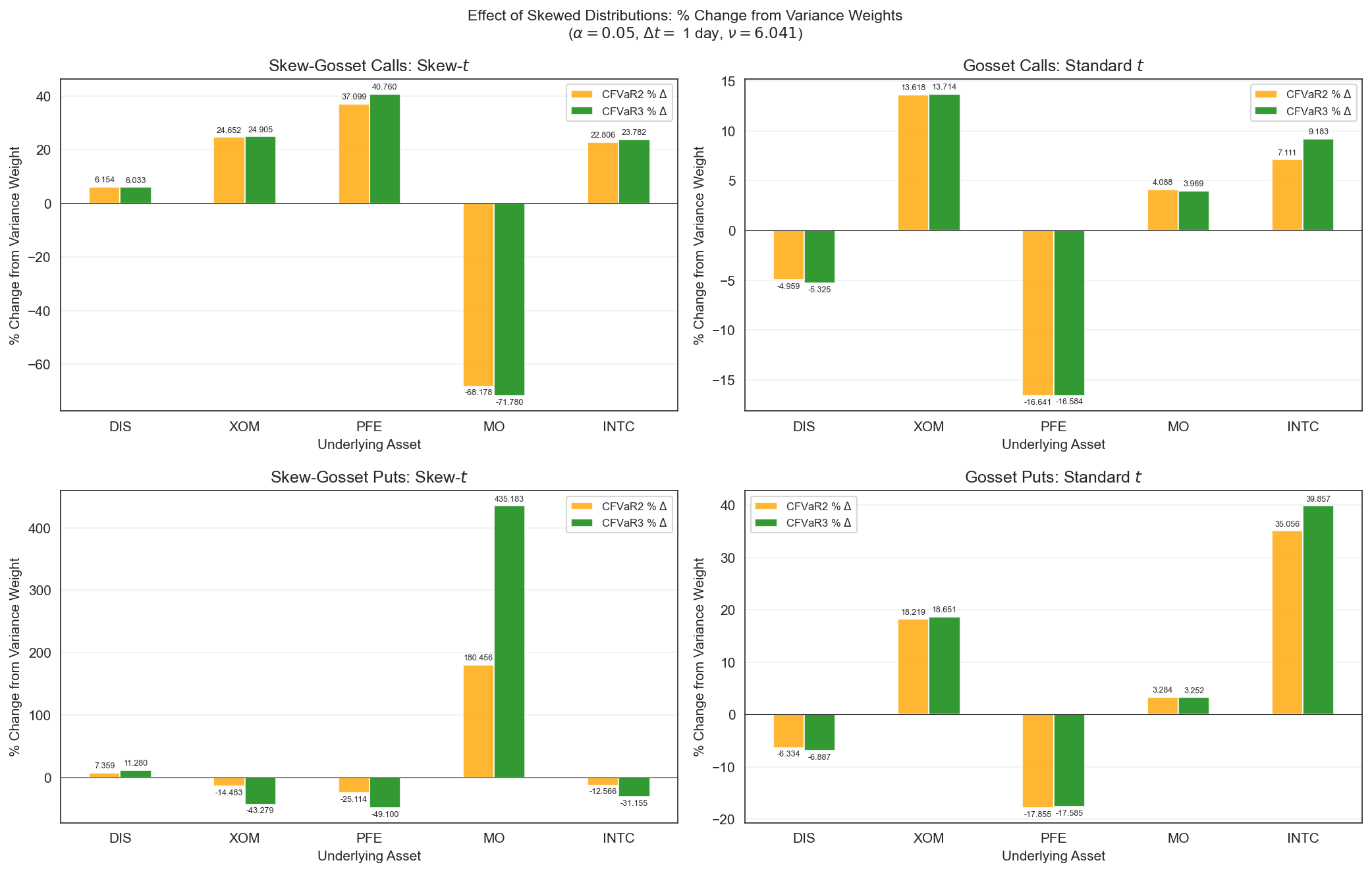}
    \caption{Percentage change of portfolio weights for $\CFVaR_2$ and $\CFVaR_3$ optimal weights when compared against variance weights. We consider both the skew $t$ (left) and the standard $t$ (right) settings, and both call (top) and put (bottom) portfolios.}
    \label{fig:pct_change}
\end{figure}

The goal here is to compare the $\CFVaR_3$ and $\CFVaR_2$ optimal weights against variance.
In this subsection, we plot the optimal weights percentage change of $\CFVaR_2$ and $\CFVaR_3$ using the optimal variance weights as a benchmark. Figure~\ref{fig:pct_change} displays the results.

The $\CFVaR_3$ weights appear to depart farther from the variance than the $\CFVaR_2$ weights. For example, in the case of put options on Altria Group (MO), the optimal $\CFVaR_2$ weight is $180.456\%$ larger than the optimal variance weight, whereas the optimal $\CFVaR_3$ weight is $435.183\%$ larger than the optimal variance weight.

\section{Conclusion} \label{sec:conclusion}

In this paper, we analyze optimal option portfolios under variance and VaR minimization, assuming that the underlying asset returns are skew $t$-distributed. We approximate the VaR with the Cornish-Fisher expansion, and in the case of two-terms, we find a closed-form expression for the optimal number of shares. We then use numerical optimization to compute the optimal portfolio for the case of the three-term Cornish-Fisher expansion. Finally, we showcase the optimal portfolios under three risk measures and compare the results with the standard $t$-distribution.

Our findings show that the optimal weights for the three-term expansion only departs more than the two-term-expansion from the variance weights in the case of the skew $t$ distribution, showing that robust portfolios require consideration of skew distributions; this effect does not persist in the case of the standard $t$ distribution. Further, the portfolio outlook may change in the standard $t$-distribution.

In future work, we aim to extend our results to include trading costs as in \citet{AlexanderS.2006MCaV}, more periods as done by~\citet{Deng2019}, or including the effect of liquidity on the asset prices as considered by~\citet{Shidfara2014,Yazdanian2016}.

\paragraph{Acknowledgements} Traian A. Pirvu acknowledges that this work was
supported by NSERC grant RGPIN-2019-05397.

\bibliographystyle{abbrvnat}
\bibliography{references}

\appendix

\section{Skew-Elliptical $t$-Distribution}
We now characterize the skew $t$-distribution \citep{AzzaliniCapitanio2003Skewt}.
Fix the following parameters: a degrees of freedom $\nu >0$, location vector $\MU \in \R^N$, skewness vector $\OMEGA \in \R^N$, and scale matrix $\SIGMA \in \R^{N\times N}$. Say a random vector $\Y \in \R^N$ has the \textit{skew} $t$-distribution and write \[
    \Y \sim t_N^{\operatorname{skew}} (\MU, \SIGMA, \OMEGA, \nu)
\]
if $\Y$ has p.d.f.\
\[
    f_\Y(\y) = 2t_N (\y; \nu) T_1
    \left\{
        \OMEGA^\top (\y - \MU) 
        \sqrt{
            \dfrac{\nu + N}{Q_\y + \nu}
        }
        ;
        \nu + N
    \right\},
\]
where $T_1(\,\cdot\,; \nu + N)$ is the p.d.f.\ of a scalar $t$-distribution with $\nu + N$ degrees of freedom and where $t_N(\,\cdot\,; \nu)$ denotes the p.d.f.\ of an $N$-dimensional $t$-distribution with $\nu$ degrees of freedom, given by:
\[
    \begin{split}
        Q_\y &\coloneqq (\y - \MU)^\top \SIGMA^{-1} (\y- \MU),
        \\
        t_N(\y; \nu) &\coloneqq \dfrac{\GammaFunc(\tfrac{\nu + N}{2})}{\sqrt{\det(\SIGMA)} (\pi\nu)^{N/2} \GammaFunc(\nu/2)} (1 + Q_\y /\nu)^{-\tfrac{\nu + N}{2}}
        ,
    \end{split}
\]  
and $\operatorname{Gamma}(\,\cdot \,)$ denotes the Gamma function. 

\begin{remark}
    For $\OMEGA = \Zero$, the skew-elliptical $t$ p.d.f.\ reduces to the p.d.f.\ of the standard $t$-distribution $t_N(\MU, \SIGMA, \nu)$.
\end{remark}

\citet{KIM2003417} provide explicit expressions for the moments of quadratic forms of skew-elliptical random variables, which we make use of.

\section{Proof of Theorem~\ref{theorem:cfvar2_optim_t_dist_returns_nonzero_drift}} \label{sec:proofs}

\begin{proof} 
    We proceed with a modification to Lagrange multipliers by setting the linear term to a constant $\varepsilon$, minimizing the radical term, then minimizing over all possible $\varepsilon$.

    Setting $\varepsilon \coloneqq \mbf u^\top \x$, we have the optimization problem 
    \[
        \begin{cases}
            \min \limits_{\x \in \R^M} \left\{ 
            -\varepsilon - \normal^{-1}(\alpha) \sqrt{\dfrac{1}{2} \x^\top Q \x}
            \right\}
            \\[12pt]
            \v (\S, t)^\top \x = 1
            \\[8pt]
            \mbf u^\top \x = \varepsilon
            .
        \end{cases}
    \]
    We can rewrite the constraints using $J$ as in Equation~\eqref{eqn:J_psistar} and with  
    $
        \PSI \coloneqq \begin{bmatrix}
            \varepsilon
            \\
            1
        \end{bmatrix}
    $.
    Thus we can express the optimization problem as
    \[
        \begin{cases}
            \min \limits_{\x \in \R^M} \left\{ 
            -\varepsilon - \normal^{-1}(\alpha) \sqrt{\dfrac{1}{2} \x^\top Q \x}
            \right\}
            \\[12pt]
            J \x = \PSI
            .
        \end{cases}
    \]
    Assuming that $\alpha < 1/2$, we have $\normal^{-1}(\alpha) < 0$, so by monotonicity of the square root, the objective function can be written as
    \[
        \begin{cases}
            \min \limits_{\x \in \R^M} \left\{ 
            \dfrac{1}{2} \x^\top Q \x
            \right\}
            \\[12pt]
            J \x = \PSI
            .
        \end{cases}
    \]
    This quadratic program can be easily solved to obtain the following:
    \begin{equation*}
        \x = Q^{-1} (J^\top \BETA),
    \end{equation*}
    where $\BETA$ is the vector of Lagrange multipliers. Substituting this into the constraints yields
    \begin{equation*}
        \BETA
        =
        (J Q^{-1} J^\top )^{-1} \PSI
        \quad\text{and}\quad
        \x 
        = 
            Q^{-1} 
            \left(
                J^\top (J Q^{-1} J^\top )^{-1} \PSI
            \right)
            .
    \end{equation*}
    Next, set
    \[
        G 
        \coloneqq
            Q^{-1} J^\top (J Q^{-1} J^\top )^{-1}
        = 
        \begin{bmatrix}
            g_{1,1} & g_{1,2}
            \\
            \vdots & \vdots
            \\
            g_{N,1} & g_{N,2}
        \end{bmatrix}
        ,
    \]
    and thus we may write
    \[
        \x 
        = 
        G \PSI
        .
    \]
    This produces an expression for the variance term in the objective function in terms of $\varepsilon$:\footnote{This approach is also used by \citet{Pan2023}.}
    \begin{align*}
        \dfrac{1}{2} \x^\top Q \x
        &=
            \Ascr \varepsilon^2 
            + 
            \Bscr \varepsilon 
            + 
            \Cscr
        ,
    \end{align*}
    where
    \begin{align*}
        \Ascr &= \frac{1}{2} (G_{[\bullet,1]})^\top Q G_{[\bullet, 1]}
        \\
        \Bscr &= (G_{[\bullet,2]})^\top Q G_{[\bullet, 1]}
        \\
        \Cscr &= \frac{1}{2} (G_{[\bullet,2]})^\top Q G_{[\bullet, 2]}.
    \end{align*}
    For any given $\varepsilon$, these values yield:
    \begin{equation}
        \label{eqn:cfvar2_as_function_of_eps}
        \CFVaR_2^\alpha [\Delta V(\x)]
        =
        -\varepsilon - \normal^{-1}(\alpha) 
        \sqrt{
            \Ascr \varepsilon^2 + \Bscr \varepsilon + \Cscr
        }
        .
    \end{equation}
    Let
    \begin{equation*}
        \begin{split}
            \Ac &= 4\Ascr^2 (\normal^{-1}(\alpha))^2 - 4\Ascr
            \\
            \Bc &= 4\Ascr\Bscr(\normal^{-1}(\alpha))^2 - 4\Bscr
            \\
            \Cc &=\Bscr^2 (\normal^{-1}(\alpha))^2 - 4\Cscr
            \\
            \varepsilon_\pm &= \frac{-\Bc \pm \sqrt{\Bc^2 - 4\Ac\Cc}}{2\Ac}
        .
        \end{split}
    \end{equation*}
    The right-hand side of Equation~\eqref{eqn:cfvar2_as_function_of_eps} is minimized by \[
        \varepsilon^\star 
        \coloneqq 
        \begin{cases}
            \varepsilon_+ & \text{if~} 2\Ascr \varepsilon_+ + \Bscr > 0
            \\
            \varepsilon_- & \text{if~} 2\Ascr \varepsilon_- + \Bscr > 0.
        \end{cases}
        \qedhere
    \]
\end{proof}

\section{Hu and Kercheval Student $t$ Parameter Estimation}

\begin{table}[H]
    \centering
    \caption{$t$-Distributed Underlying Returns~\citep[Table 4]{Hu01012010};~\citep[Table 3.1 and 3.2]{Pan2023}}
    \begin{tabular}{lccccc}
        \toprule
        4/8/2005 & Disney & Exxon & Pfizer & Altria & Intel \\
        \midrule
        Stock Price ($S$) & 28.02 & 60.01 & 25.24 & 65.53 & 23.29 \\
        Annualized Expected Return ($\mu$) & 0.0151 &  0.0800 & $-0.0178$ &  0.0714 &  0.0305 \\
        Annualized Volatility ($\sigma$) & 0.1699 & 0.2032 & 0.2064 & 0.1794 & 0.2476 \\
        \midrule
        Correlations: Disney & 1     &       &       &       &       \\
        Exxon                 & 0.363 & 1     &       &       &       \\
        Pfizer                & 0.378 & 0.373 & 1     &       &       \\
        Altria                & 0.265 & 0.271 & 0.259 & 1     &       \\
        Intel                 & 0.460 & 0.324 & 0.349 & 0.225 & 1     \\
        \bottomrule
    \end{tabular}
    \label{table:data}
\end{table}

\end{document}